\documentclass[prb,twocolumn,amsfonts,showpacs]{revtex4}
\usepackage{graphics,bm,natbib}
\begin{document}
\widetext
\title{Lattice Distortion and Magnetic Ground State of YTiO$_3$ and LaTiO$_3$}
\author{I. V. Solovyev}
\email[Electronic address: ]{igor@issp.u-tokyo.ac.jp}
\affiliation{
PRESTO-JST, Institute for Solid State Physics, University of Tokyo,\\
Kashiwanoha 5-1-5, Kashiwa, Chiba 277-8531, Japan
}
\date{\today}

\widetext
\begin{abstract}
Effects of lattice distortion on the magnetic ground state of
YTiO$_3$ and LaiO$_3$ are investigated on the basis accurate tight-binding
parametrization of the $t_{2g}$ electronic structure extracted from the local-density
approximation. The complexity of these compounds is related with the fact that
the $t_{2g}$-level splitting, caused by lattice distortions, is comparable with
the energies of superexchange and spin-orbit interactions. Therefore, all these interactions are
equally important and should be treated on an equal footing. The Hartree-Fock
approximation fails to provide a coherent description simultaneously for
YTiO$_3$ and LaTiO$_3$, and it is essential to go beyond.
\end{abstract}



\maketitle

  Among the large variety of transition-metal perovskite oxides, YTiO$_3$ (YTO) and LaTiO$_3$ (LTO)
have received a particular attention. Both are regarded as prototypical examples of Mott-Hubbard
insulators. It appears, however, that these, formally isoelectronic compounds (having one
$3d$ electron in the triply-degenerate $t_{2g}$ shell),
exhibit very different magnetic properties: YTO is a
ferromagnet, whereas LTO is a three-dimensional (G-type) antiferromagnet.
Another puzzling feature is the nearly isotropic magnon spectrum, observed both
in YTO and LTO despite a noticeable orthorhombic distortion.\cite{Keimer,Ulrich}

  Owing to the fractional population of the $t_{2g}$
manifold, the orbital degrees of freedom are expected to play a very important role
and affect the magnetic properties.
However, the theories proposed in this context crucially
depend on several factors, and there are two points of view which are currently
discussed in the literature. (i) The first one is based on the generalization of the
superexchange (SE) theory of spin and orbital interactions between \textit{degenerate} $t_{2g}$
levels. It starts with the spin-orbital SE model
by Kugel and Khomskii (K\&K),\cite{KugelKhomskii}
and exploits
the idea of
orbital fluctuations, which are inherent to this model.\cite{Khaliullin}
(ii) The spin and orbital structure is
fully determined by lattice distortions,
which \textit{lift the orbital degeneracy}.\cite{FujitaniAsano,SawadaTerakura,MochizukiImada_JPS,MochizukiImada_PRL,Cwik}
The role of (relativistic) spin-orbit (SO) interaction has been also emphasized.\cite{Mizokawa}

  Therefore, there are two important questions, which can be clarified on the basis
of electronic structure calculations. (i) What is the effect of the lattice distortion on the
electronic structure of YTO and LTO? Particularly, how do the $t_{2g}$ levels split by
this distortion? (ii) What is the hierarchy between the $t_{2g}$-level splitting, the SE interaction
energy, and the SO coupling?

  The SE interaction in the bond ${\bf i}$-${\bf j}$
is basically the kinetic energy gain, which is acquired by the $t_{2g}$ electron
occupying the atomic orbital $| {\bf i} \rangle$ at the site ${\bf i}$
in the process of
virtual hoppings into the
subspace of unoccupied orbitals $\widehat{P}_{\bf j}$ at the (neighboring) site ${\bf j}$,
and vice versa:~\cite{PWA}
\begin{equation}
\varepsilon_{\bf ij}^{\sigma \sigma'} =
-\frac{\alpha_{\bf ij}^{\sigma \sigma'}}{\Delta E^{\sigma \sigma'}} \equiv
- \frac{\langle {\bf i} | \widehat{t}_{\bf ij} \widehat{P}_{\bf j}
\widehat{t}_{\bf ji} | {\bf i} \rangle  + ({\bf i} \leftrightarrow {\bf j}) }{\Delta E^{\sigma \sigma'}},
\label{eqn:energygain}
\end{equation}
where
$\sigma$ and $\sigma'$ are the
spin states associated with the sites ${\bf i}$ and ${\bf j}$, respectively, and
the transfer interactions $\widehat{t}_{\bf ij}$ are allowed only between orbitals with the
same spin.
For the nearest-neighbor interactions in the perovskite lattice,
it is sufficient
to consider two collinear configurations,
$\sigma \sigma'$$=$$\uparrow \uparrow$ and $\uparrow \downarrow$,
and select the ones which minimize the total energy gain
$\varepsilon_{\rm T}$$=$$\frac{1}{2} \sum_{\bf ij} \varepsilon_{\bf ij}^{\sigma \sigma'}$.
In the case of the antiferromagnetic (AFM) alignment,
$\sigma \sigma'$$=$$\uparrow \downarrow$,
all orbitals with the spin $\uparrow$ at the
site ${\bf j}$
are located in the unoccupied
part of the spectrum and available for the hoppings. Therefore,
$\widehat{P}_{\bf j}$$=$$1$ and
$\alpha_{\bf ij}^{\uparrow \downarrow}$$=$$\langle {\bf i} | \widehat{t}_{\bf ij} \widehat{t}_{\bf ji} |
{\bf i} \rangle$$+$$({\bf i}$$\leftrightarrow$${\bf j})$.
In the ferromagnetic (FM or F) case, $\sigma \sigma'$$=$$\uparrow \uparrow$,
the occupied orbital $| {\bf j} \rangle$ should be
excluded from the subspace $\widehat{P}_{\bf j}$. This yields
$\widehat{P}_{\bf j}$$=$$1$$-$$| {\bf j} \rangle \langle {\bf j} |$ and
$\alpha_{\bf ij}^{\uparrow \uparrow}$$=$$\alpha_{\bf ij}^{\uparrow \downarrow}$$-$$\Delta\alpha_{\bf ij}$, where
$\Delta\alpha_{\bf ij}$$=$$2 \left| \langle {\bf i} | \widehat{t}_{\bf ij} | {\bf j} \rangle \right|^2$.
$\Delta E^{\sigma \sigma'}$ is the on-site Coulomb interaction
between two $3d$ electrons, which also depends on the spin state: $\Delta E^{\uparrow \downarrow}$$=$$U$ while
$\Delta E^{\uparrow \uparrow}$$=$$U$$-$$J$, where
$U$ is the Coulomb repulsion and $J$ is the intra-atomic exchange coupling.\cite{comment.1}
Because of this $J$, the ''orthogonal'' orbitals,
which do not interact via the kinetic energy term,
$\langle {\bf i} | \widehat{t}_{\bf ij} | {\bf j} \rangle$$=$$0$,
tend to stabilize the FM structure.
In the opposite limit $\Delta \alpha_{\bf ij}$$\simeq$$\alpha_{\bf ij}^{\uparrow \downarrow}$,
the FM alignment does not lead to any energy gain, and
the coupling will be AFM.

  The alternation of occupied orbitals at different atomic sites (the orbital ordering -- OO)
should be found
variationally and minimize $\varepsilon_{\rm T}$.
This is the basic idea of the K\&K theory.\cite{KugelKhomskii}
The orbital interactions have the same origin as the spin SE.
Therefore, the energy gain associated with the OO
is of the order of
$\varepsilon_{\rm T}$$\sim$$1/U$, and there is a strong interplay
between spin and orbital degrees of freedom.
In the degenerate case, one can always find some orthogonal configuration of the occupied
orbitals, which in the
single-determinant Hartree-Fock (HF) approach corresponds to the FM
ground state (GS).
However, the HF solutions
remain degenerate with respect to some
number of orbital
configurations.
This degeneracy leaves a room for orbital fluctuations, which may
alter the HF conclusion about the form of the magnetic GS.\cite{Khaliullin}

  An alternative mechanism of the OO is the lattice distortions, which
lifts the orbital degeneracy and
acts as an external
field constraining the form of occupied orbitals in Eq.~(\ref{eqn:energygain}).
Since the orbital degeneracy is lifted, the HF approach may be justified.\cite{SawadaTerakura,MochizukiImada_PRL}
This mechanism is proportional to the electron-phonon coupling, and will
dominate over the K\&K SE mechanism in the large-$U$ limit.
Then, the OO does not depend on the magnetic state
and the mapping onto the Heisenberg model yields the following expression for
$J_{\bf ij}$$=$$\frac{1}{2}(\varepsilon_{\bf ij}^{\uparrow \downarrow}$$-$$
\varepsilon_{\bf ij}^{\uparrow \uparrow})$:
\begin{equation}
J_{\bf ij}=\frac{\alpha_{\bf ij}^{\uparrow \downarrow}}{2}
\frac{J/U - \Delta \alpha_{\bf ij}/\alpha_{\bf ij}^{\uparrow \downarrow}}{U-J},
\label{eqn:Jdistortion}
\end{equation}
which can be both FM and AFM, depending on the ratio of $J/U$ and
$\Delta \alpha_{\bf ij}/\alpha_{\bf ij}^{\uparrow \downarrow}$.

\begin{figure}[t!]
\begin{center}
\resizebox{3.8cm}{!}{\includegraphics{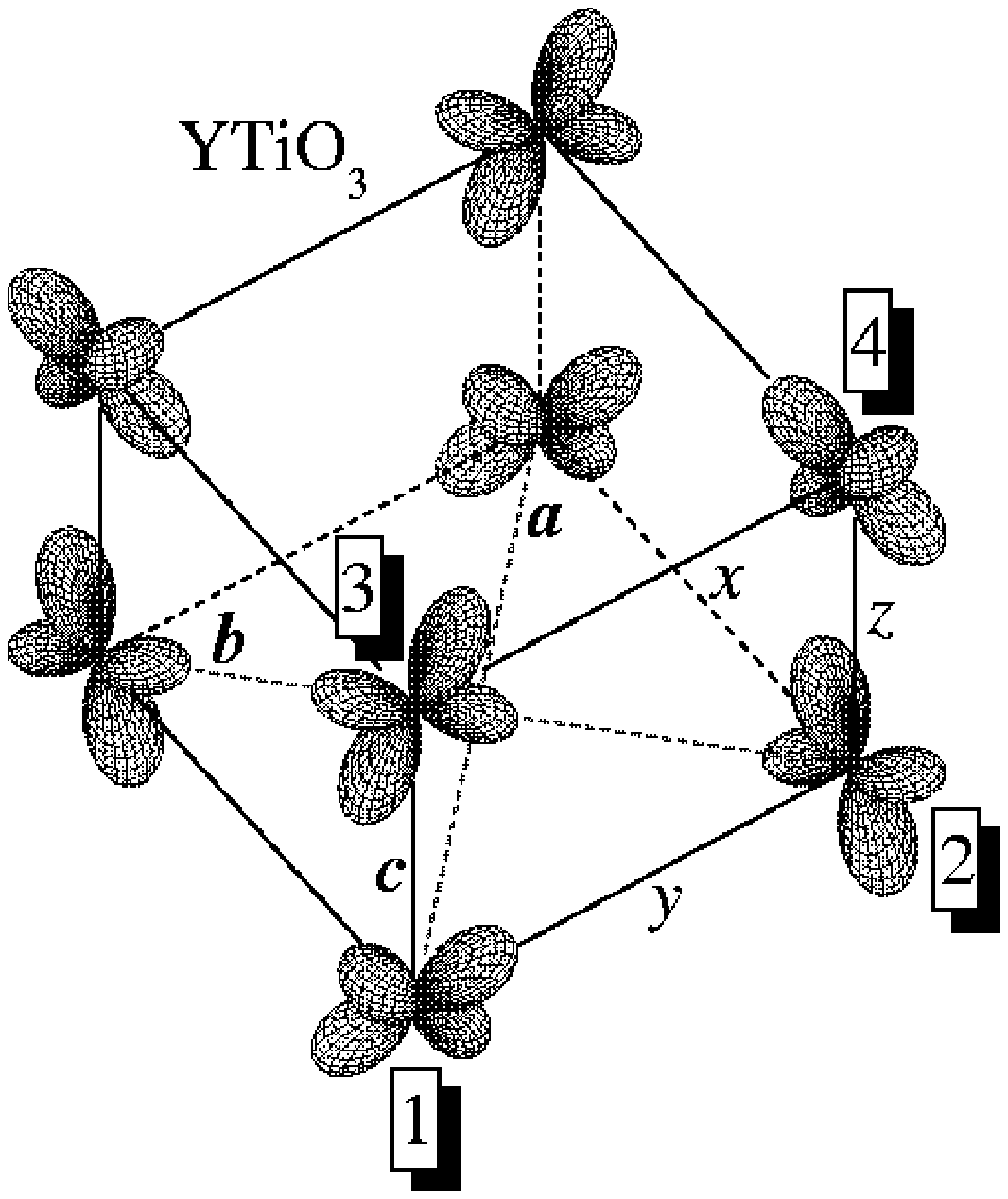}}
\resizebox{3.8cm}{!}{\includegraphics{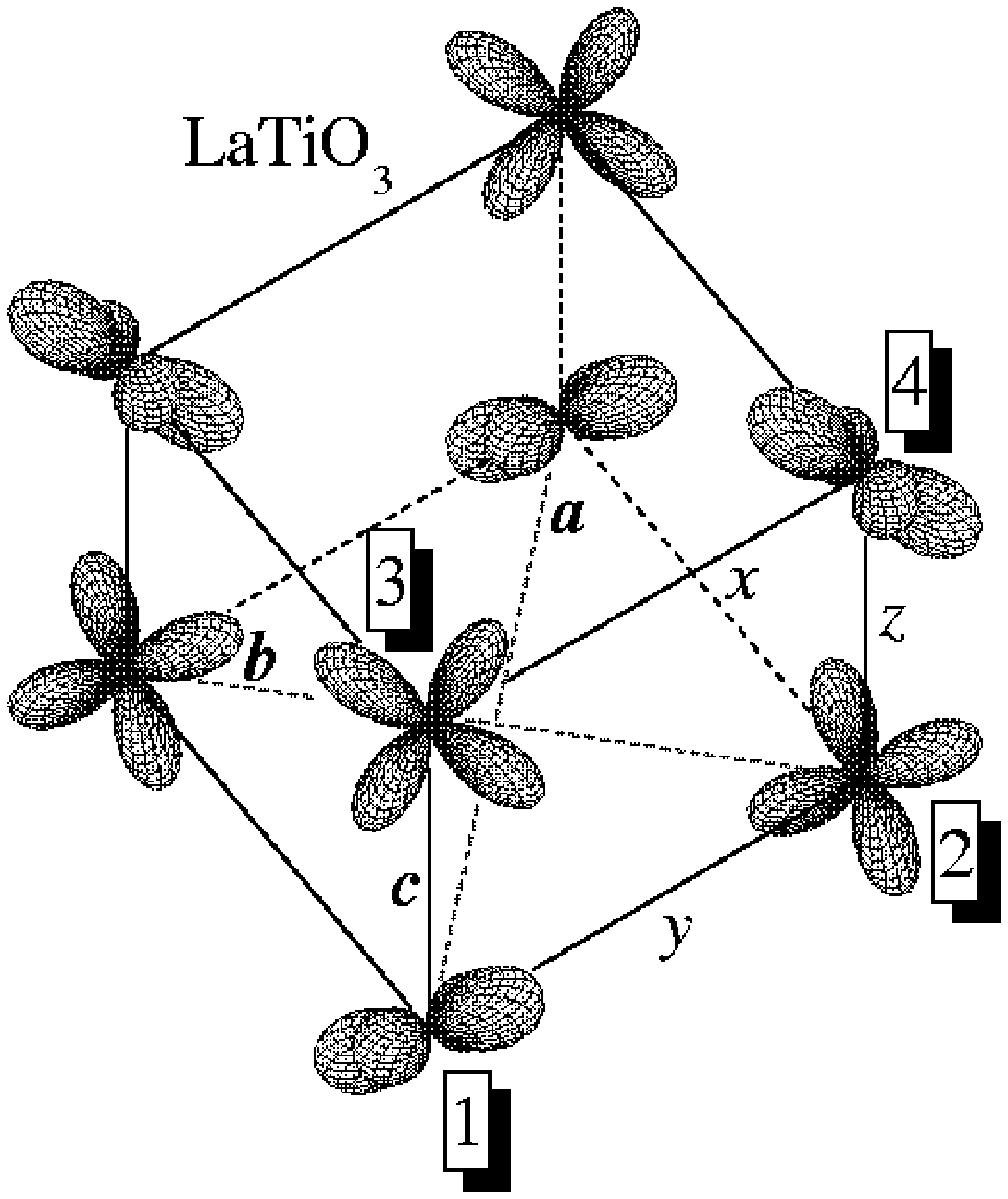}}
\end{center}
\caption{\label{fig.OO1} $t_{2g}$-electron densities
obtained in Hartree-Fock calculations
after including the spin-orbit interaction.
$x$, $y$, and $z$ are the cubic axes. ${\bf a}$, ${\bf b}$, and ${\bf c}$
are the orthorhombic axes.}
\end{figure}

  Let us consider the second scenario and assume that
all relevant interactions can be described in the basis of some
local $t_{2g}$
orbitals $|X \rangle$, $|Y \rangle$,
and $|Z \rangle$, associated with the Ti sites.
Then, the occupied orbital at the site $1$
(see Fig.~\ref{fig.OO1})
can be searched in the form
$|1 \rangle$$=$$ \sin \theta \cos \phi |X \rangle$$+$$
\sin \theta \sin \phi |Y \rangle$$+$$\cos \theta |Z \rangle$,
and the ones at the sites $2$ and $3$ are automatically generated from
$|1 \rangle$ using the symmetry operations of the $D_{2h}^{16}$ group
(the $180^\circ$ rotations around the orthorhombic ${\bf a}$ and
${\bf c}$ axes, respectively).
In principle, $\theta$ and $\phi$ are uniquely determined
by the lattice distortion.

  However,
it is sometimes tempted to approach the problem from the opposite side,\cite{MochizukiImada_JPS}
and find $\theta$ and $\phi$
from the condition
$J_{12}$$=$$J_{13}$, suggested by
recent neutron scattering studies.\cite{Keimer,Ulrich}
In order to illustrate this idea, let us consider a simplified
model and choose $|X \rangle$, $|Y \rangle$, and $|Z \rangle$
as
$|yz \rangle$,
$|zx \rangle$, and $|xy \rangle$, respectively,
in the cubic coordinate frame shown in Fig.~\ref{fig.OO1}.
The transfer interactions are parameterized according to Slater and
Koster (S\&K): i.e., the only nonvanishing matrix elements along $z$
are $t_{13}^{XX}$$=$$t_{13}^{YY}$$=$$t$, etc.
Then,
it is easy to verify that the condition $J_{12}$$=$$J_{13}$ leads to
the following OO:\cite{comment.2}
$|1 \rangle$$=$$|2 \rangle$$=$$\frac{1}{\sqrt{3}}(|xy \rangle$$+$$|yz \rangle$$+$$|zx \rangle)$,
$|3 \rangle$$=$$|4 \rangle$$=$$\frac{1}{\sqrt{3}}(|xy \rangle$$-$$|yz \rangle$$-$$|zx \rangle)$,
which does not depend on $J/U$.
This is precisely the OO proposed in Ref.~\onlinecite{MochizukiImada_JPS}.
It is compatible with the orthorhombic $D_{2h}^{16}$ symmetry, and
corresponds to some local trigonal distortion,
caused by either oxygen or La displacements.\cite{MochizukiImada_PRL}

  This result, however, prompts several new questions.
(i) The magnetic coupling is expected to be AFM for all reasonable
values of $J/U$. Therefore, this would explain the
experimental situation in LTO, but not in YTO.
(ii) It is not clear whether this OO is compatible with the actual
experimental distortion observed in LTO.
Note that in the $D_{2h}^{16}$ group, only inversion centers coincide with the
Ti sites. Therefore, the local $t_{2g}$-level
splitting
is controlled by 5 independent parameters, which may include
both trigonal and Jahn-Teller modes. All distortions are formally equivalent,
at least from the viewpoint of $D_{2h}^{16}$ symmetry, and \textit{a priori}
there is no reason why the particular trigonal mode should dominate.
In addition to the $t_{2g}$-level
splitting, the crystal distortion may also affect
the transfer interactions
through the buckling
of the Ti-O-Ti bonds.\cite{comment.3}
(iii) What are the roles of the K\&K mechanism and the SO interaction? Are they totally
quenched by the lattice distortion, as it was suggested in
Refs.~\onlinecite{MochizukiImada_PRL,Cwik}?
The situation should be carefully checked,
and it is important to turn to
first-principles calculations, which automatically
include all these ingredients.

  We use the linear-muffin-tin-orbital (LMTO) method,\cite{LMTO} and
employ the tight-binding (TB) parametrization
of the $t_{2g}$ bands, obtained in the local-density approximation (LDA) for the
experimental crystal structures.\cite{Cwik}
The latter step is achieved through the
downfolding procedure. A similar analysis has been undertaken in Ref.~\onlinecite{Pavarini}.
(i) Each LMTO eigenvector is divided in two parts:
$| t \rangle$, which is expanded over the local $t_{2g}$ orbitals
$| X \rangle$, $| Y \rangle$, and $| Z \rangle$ at each Ti site, and
$| r \rangle$, which is expanded over the rest of the basis functions.
The corresponding secular equation, which holds for
the LMTO Hamiltonian $\widehat{\cal H}$, is given by
\begin{eqnarray}
( \widehat{\cal H}_{tt}-E ) | t \rangle  +  \widehat{\cal H}_{tr} | r \rangle & = & 0, \label{eqn:seceq1}\\
\widehat{\cal H}_{rt} | t \rangle  +  ( \widehat{\cal H}_{rr}-E ) | r \rangle & = & 0. \label{eqn:seceq2}
\end{eqnarray}
(ii) By eliminating $| r \rangle$ from Eq.~(\ref{eqn:seceq2}) one obtains an effective
$E$-dependent Hamiltonian:
$\widehat{\cal H}_{tt}^{\rm eff}(E)$$=$$\widehat{\cal H}_{tt}$$-$$\widehat{\cal H}_{tr}
(\widehat{\cal H}_{rr}$$-$$E)^{-1}\widehat{\cal H}_{rt}$, where
$| t \rangle$ obeys the condition
$\langle t | \widehat{S} | t \rangle$$=$$1$ and
$\widehat{S}(E)$$=$$1$$+$$\widehat{\cal H}_{tr}
(\widehat{\cal H}_{rr}$$-$$E)^{-2}\widehat{\cal H}_{rt}$.
(iii) The TB parameters $\widehat{t}$$\equiv$$ \| \widehat{t}_{\bf ij} \|$
are obtained after the orthonormalization of the vectors
$| t \rangle$$\rightarrow$$| \tilde{t} \rangle$$=$$\widehat{S}^{1/2}| t \rangle$:
\begin{equation}
\widehat{t}(E) = \widehat{S}^{-1/2}(E) \widehat{\cal H}_{tt}^{\rm eff}(E) \widehat{S}^{-1/2}(E),
\label{eqn:TB}
\end{equation}
Finally, $E$ is fixed to the center of the
$t_{2g}$ band.

\begin{figure}[t!]
\begin{center}
\resizebox{4.0cm}{!}{\includegraphics{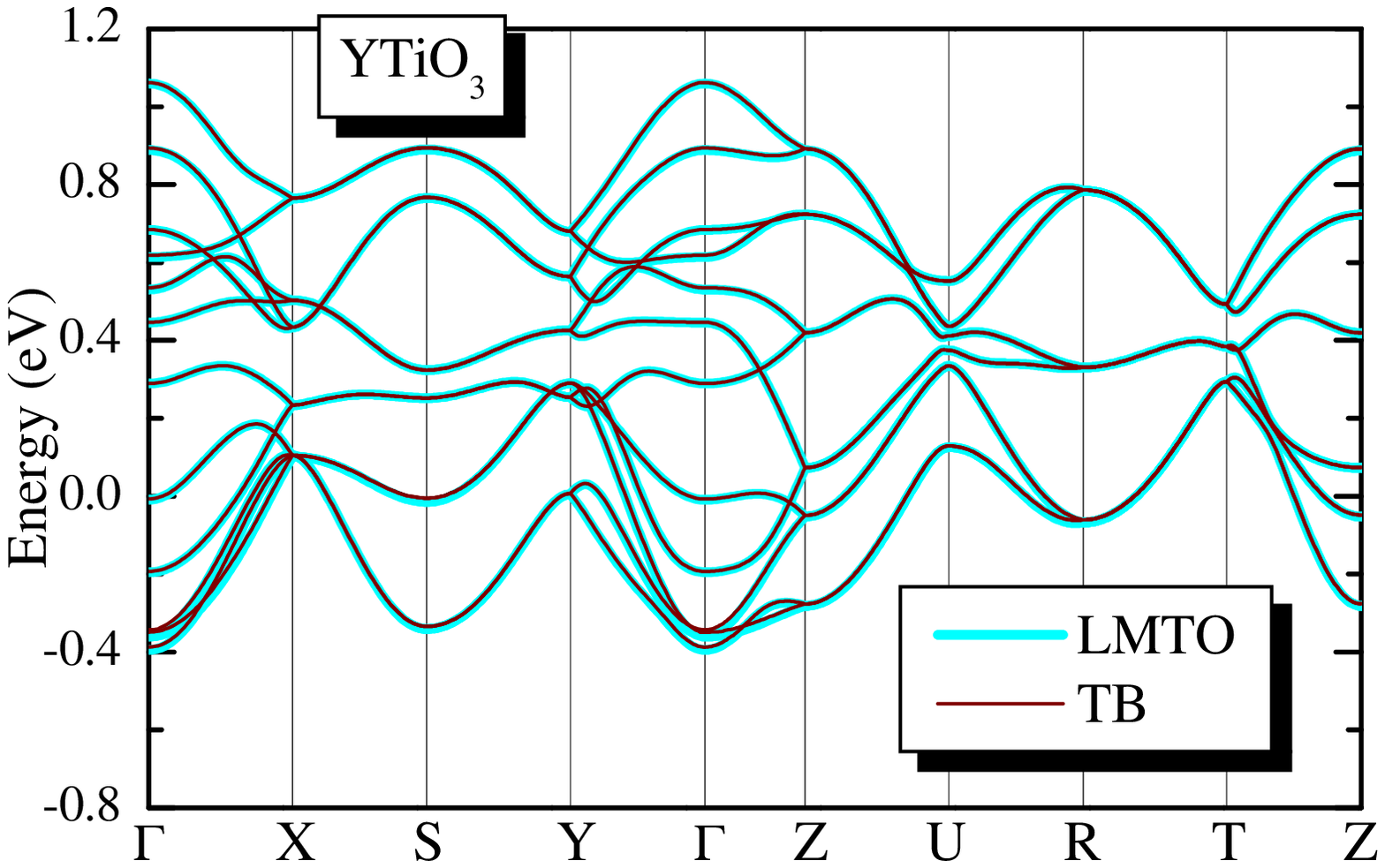}}
\resizebox{4.0cm}{!}{\includegraphics{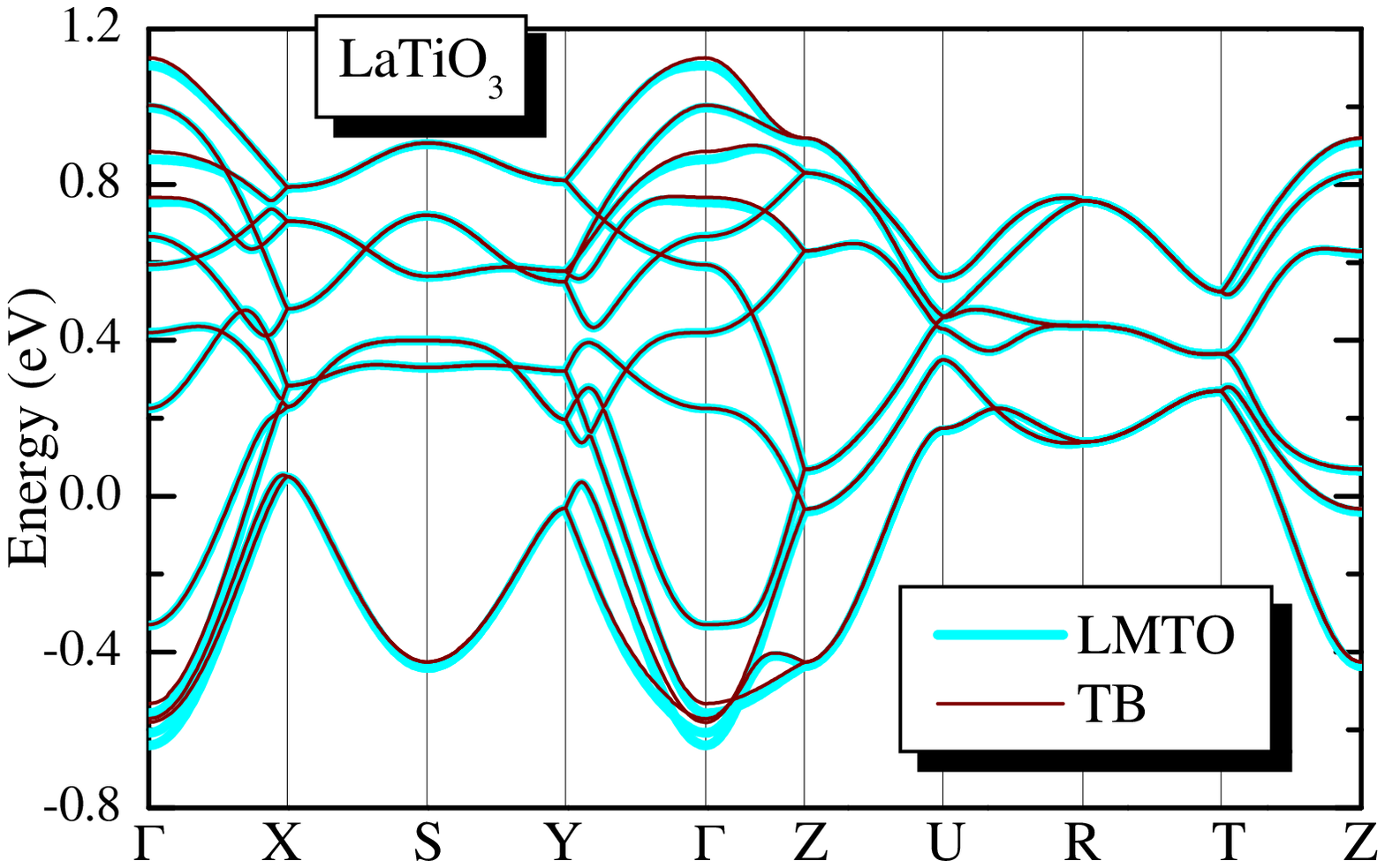}}
\end{center}
\caption{\label{fig.bands} Energy bands obtained in LMTO calculations and
after tight-binding (TB) parametrization.}
\end{figure}

  The choice of the local $t_{2g}$ orbitals is somewhat ambiguous.
In our case we first calculated the site-diagonal elements of the density matrix
in the basis of \textit{all} Ti($3d$) orbitals and taking into account
the contributions of only the $t_{2g}$ bands shown in Fig.~\ref{fig.bands}. This yields the
$5$$\times$$5$ matrices at each Ti site. Then,
we assign
three most populated orbitals obtained after
the diagonalization of these matrices to
$| X \rangle$, $| Y \rangle$, and $| Z \rangle$.

  The mapping onto the TB model is nearly perfect and well reproduces the behavior
of LMTO bands (Fig.~\ref{fig.bands}). Then, the site-diagonal elements of
$t_{\bf ij}$ describe the crystal-field (CF) splitting caused by lattice distortions,
and the off-diagonal elements
have a meaning of transfer interactions.
Thus, we are ready to calculate the SE interactions in the
strong coupling limit, assuming that the form of occupied orbitals is solely
determined by $t_{\bf ii}$, and using these orbitals in subsequent calculations of
$\alpha_{\bf ij}^{\uparrow \downarrow}$ and $\Delta \alpha_{\bf ij}$.
The results are summarized in Table~\ref{tab:SE}.
\begin{table}[b!]
\caption{The crystal-field (CF) splitting of the $t_{2g}$ states (in meV)
and parameters of superexchange interactions (in 10$^{-3}$eV$^2$)
in the strong-coupling limit.}
\label{tab:SE}
\begin{ruledtabular}
\begin{tabular}{cccccc}
compound  & CF splitting & $\alpha_{12}^{\uparrow \downarrow}$ & $\Delta \alpha_{12}$ &
$\alpha_{13}^{\uparrow \downarrow}$ & $\Delta \alpha_{13}$   \\
\colrule
YTiO$_3$  &  $-$$69$, $-$$42$, $112$                               & 20 &  0 & 26 & 23 \\
LaTiO$_3$ &  $-$$49$, $\phantom{-}$$\phantom{4}5$, $\phantom{1}44$ & 51 & 19 & 57 & 35 \\
\end{tabular}
\end{ruledtabular}
\end{table}

  (i) The CF splitting is larger in YTO, mainly due to the Jahn-Teller
distortion,\cite{Cwik} which is reflected in the upward shift of
one of the $t_{2g}$ levels.\cite{comment.4}
The CF splitting in LTO is not particularly strong (in fact it is considerably weaker than
the model estimates presented in Refs.~\onlinecite{MochizukiImada_PRL,Cwik}).
The inter-atomic
interactions $\alpha_{\bf ij}^{\uparrow \downarrow}$ are larger in
the less distorted LTO,
that well correlates with the larger $t_{2g}$ bandwidth
(Fig.~\ref{fig.bands}).
(ii) Both compounds exhibit certain tendency to
\textit{A-type} antiferromagnetism, which is especially strong in YTO:
since $\Delta \alpha_{12}$$\sim$$0$ and $\Delta \alpha_{13}$$\sim$$\alpha^{\uparrow \downarrow}_{13}$,
the bonds $1$-$2$ and
$1$-$3$ are expected to be FM and AFM, respectively, for all physical values of
$J/U$. Therefore, the crystal distortion alone cannot
explain the FM GS of YTO.\cite{comment.6}
The situation is somewhat milder
in LTO where the experimental G-type AFM ordering can be
stabilized for $J/U$$<$$\Delta \alpha_{12}/\alpha_{12}^{\uparrow \downarrow}$$\sim$$0.37$.\cite{comment.1}
However, even in this case
the interatomic magnetic interactions are expected to be anisotropic.
(iii) Realistic estimates for the on-site Coulomb interaction $U$ in the
$t_{2g}$ band typically vary from $3.2$ eV, suggested by constraint-LDA calculations
and taking into account the empirical screening by the $e_g$ electrons,\cite{PRB96}
to $4.4$ eV suggested by photoemission studies.\cite{Mizokawa,comment.5}
The intra-atomic exchange coupling can be estimated as $J$$\sim$$0.9$ eV.\cite{Mizokawa,PRB96,comment.5}
Therefore, $\varepsilon_{\rm T}$ can be as large as
$10$-$40$ meV per one Ti site.
This value can be used as a rough estimate for the OO stabilization
energy caused by SE interactions, which is \textit{comparable} with the CF splitting.
Therefore, the K\&K mechanism remains robust even in the distorted
perovskite compounds.
As we will see below, it may help to explain the experimentally
observed magnetic ground state in YTO (but not in LTO).
(iv) The SO interaction at the Ti sites, $\xi$$\simeq$$23$ meV, is
\textit{also comparable} with the CF splitting, and \textit{exceeds} the
total energy difference between different magnetic states
(Table~\ref{tab:J}). Therefore, it is reasonable to expect essentially noncollinear
magnetic GS with a considerable contribution of the
orbital magnetic moments.\cite{Treves}

  All these trends are clearly seen in
HF calculations, in which
the one-electron TB Hamiltonian for the $t_{2g}$ bands
was combined with the on-site Coulomb and exchange interactions extracted from the
constraint-LDA calculations
(unless it is specified otherwise),\cite{PRB96} and (optionally) the SO interaction.
The HF potential was treated in the rotationally-invariant
form.\cite{PRB03}

\begin{figure}[t!]
\begin{center}
\resizebox{3.5cm}{!}{\includegraphics{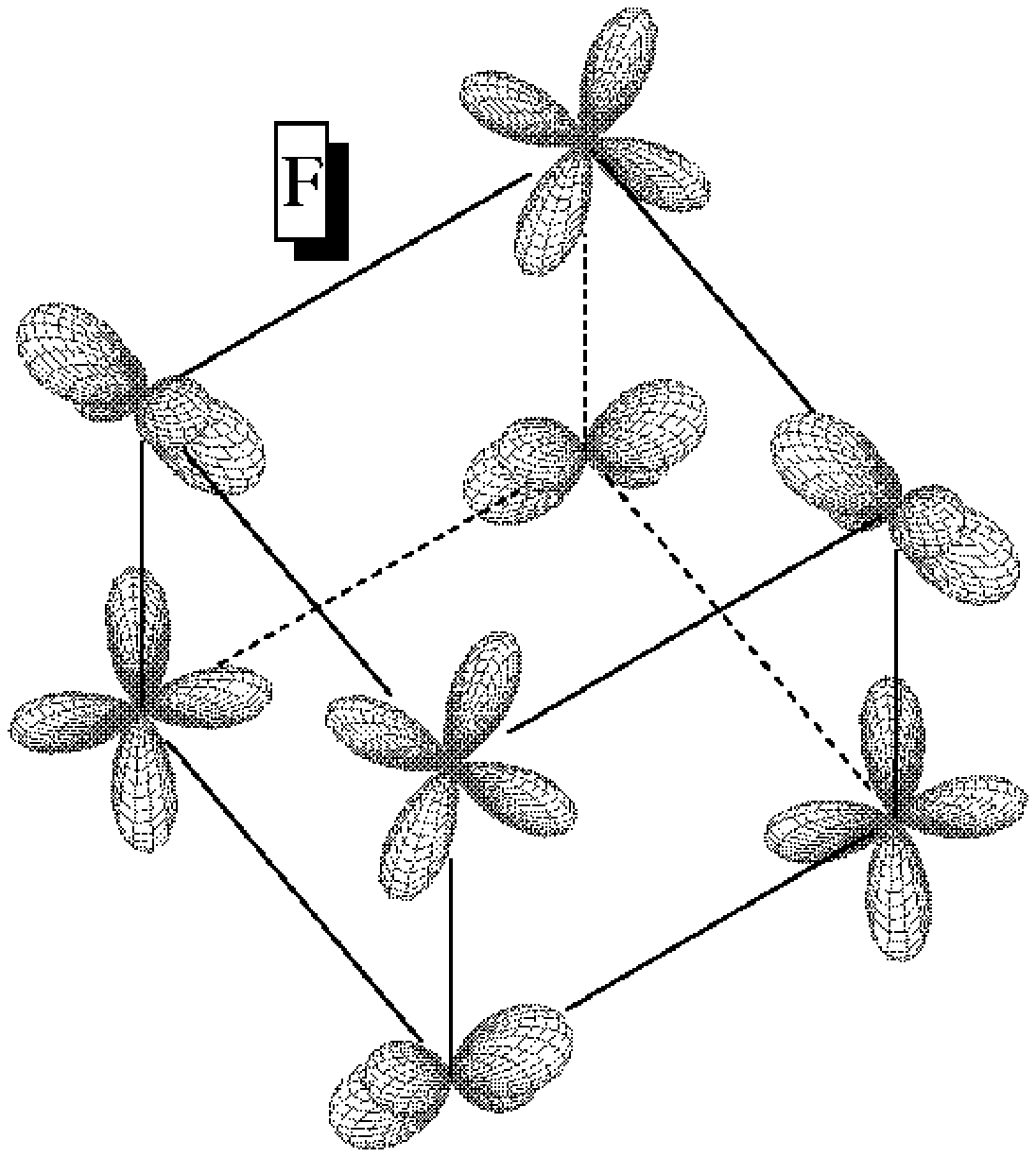}}
\resizebox{3.5cm}{!}{\includegraphics{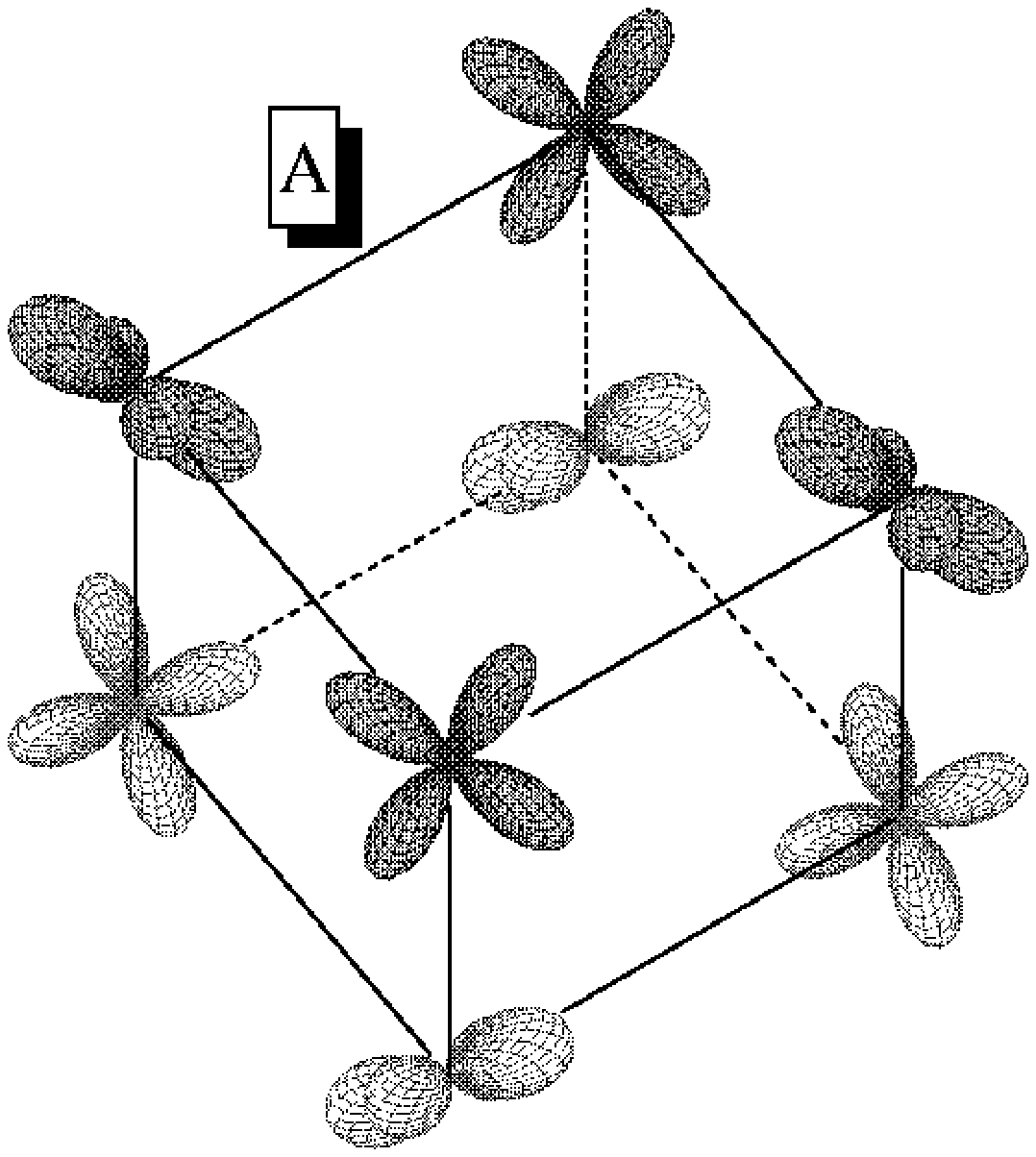}}
\\
\resizebox{3.5cm}{!}{\includegraphics{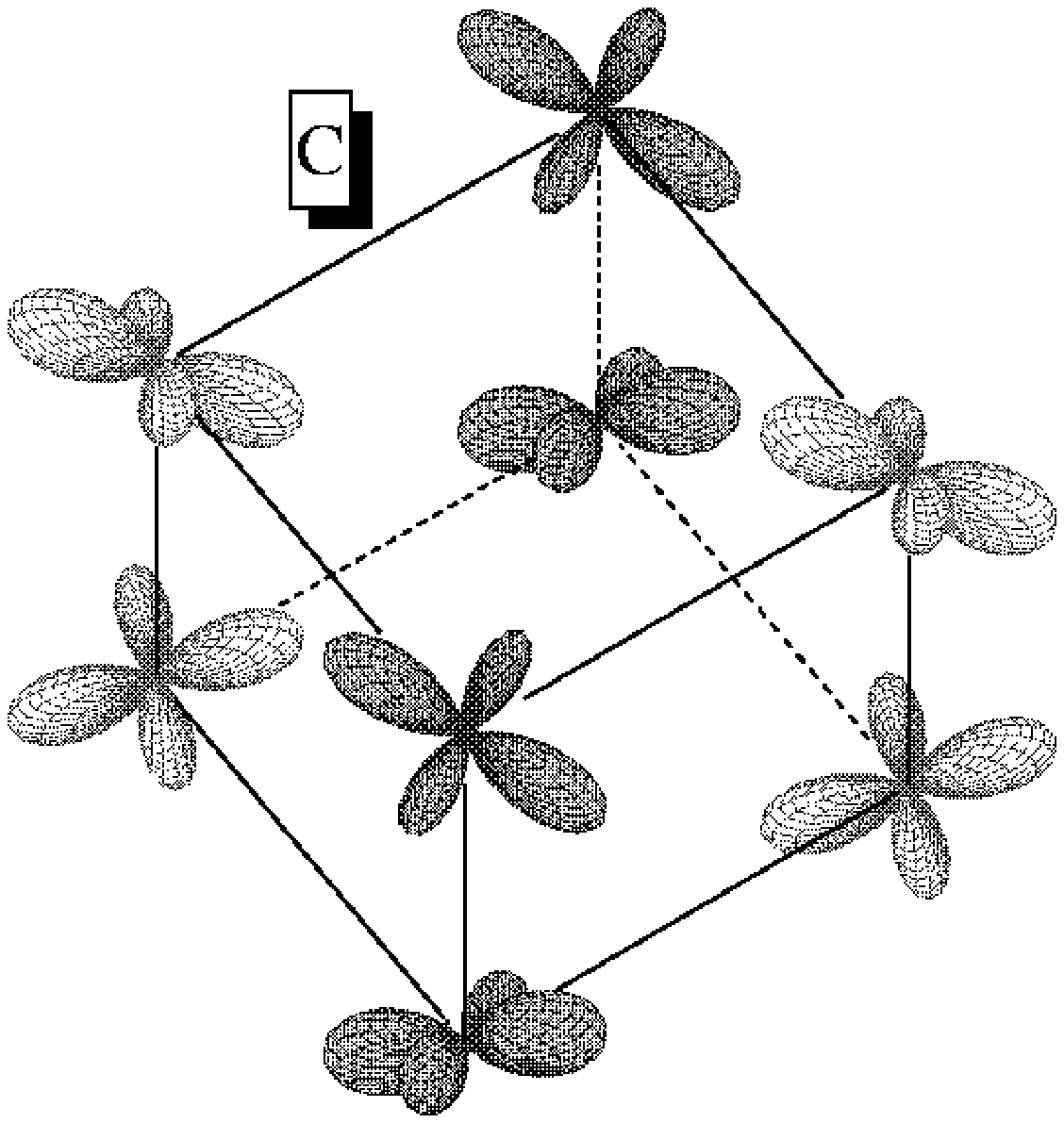}}
\resizebox{3.5cm}{!}{\includegraphics{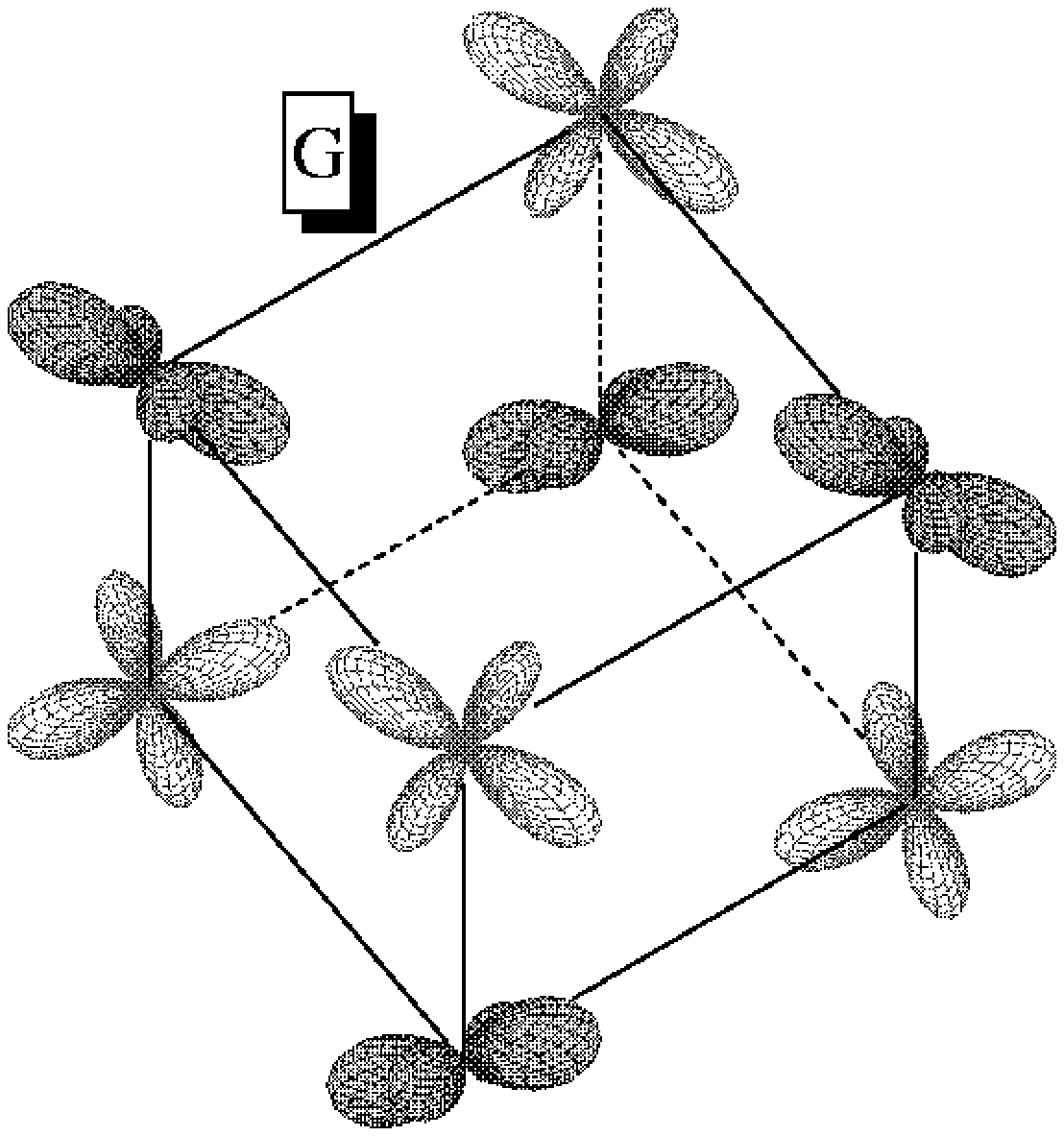}}
\end{center}
\caption{\label{fig.OO2} $t_{2g}$-electron densities in the ferromagnetic (F), A-, C-, and G-type
AFM states of LaTiO$_3$, without SO interaction.
Different spin sublattices are shown
by different colors.}
\end{figure}

  It is true that both in LTO and YTO, the OO
is strongly constrained by the lattice distortion so that
the visual change of the OO is not particularly strong in the row of FM, A-, C-, and
G-type AFM states (Fig.~\ref{fig.OO2}).
The basic question, however, is how this change is reflected in the change of other
parameters. Our main concern is the behavior of inter-atomic magnetic interactions
$J_{\bf ij}$. Since $J_{\bf ij}$ may depend on the magnetic state (through the change of
the OO), Eq.~(\ref{eqn:Jdistortion}) is no longer valid. Instead, we evaluate
$J_{\bf ij}$ separately for different magnetic states
using the
second derivatives of the total energy with respect to the angles between spin magnetic
moments.\cite{PRB03}
The results summarized in Table~\ref{tab:J}
clearly show that even
tiny change of the OO may produce a dramatic change of $J_{\bf ij}$.
In addition to the
A-type AFM ordering, expected
from the lattice distortion,
the FM state ($J_{12}$$>$$0$, $J_{13}$$>$$0$)
can be stabilized by the K\&K mechanism both in YTO and LTO.\cite{comment.7}
Since $J_{12}$$>$$0$, the G-type AFM state is unstable.
In LTO it can be stabilized only for
$U$$\sim$$4.5$ eV (which leads to $J_{12}$$=$$-$$0.3$ and $J_{13}$$=$$-$$3.4$ meV).
However, this $U$ will also destroy the FM GS in YTO
($J_{12}$$=$$-$$0.5$ and $J_{13}$$=$$-$$0.7$ meV).
Thus, there is no such parameter $U$ which would account for the experimental
behavior of both YTO and LTO on the level of mean-field HF calculations.
Contrary to the experimental finding,\cite{Keimer,Ulrich}
the magnetic interactions are strongly anisotropic.
\begin{table}[t!]
\caption{Magnetic interactions ($J$, in meV) and total energies ($E$, in meV/f.u.,
measured from the experimentally observed magnetic state) obtained in Hartree-Fock
calculations without spin-orbit interaction.}
\label{tab:J}
\begin{ruledtabular}
\begin{tabular}{ccccccc}
       &                       & YTiO$_3$              &        &                       & LaTiO$_3$          &          \\
\cline{2-7}
 phase & $J_{12}$              & $J_{13}$              & $E$    & $J_{12}$              & $J_{13}$           & $E$      \\
\hline
   F   &  $\phantom{-}$$2.0$   & $\phantom{-}$$0.6$    & $0$    & $\phantom{-}$$1.2$    & $\phantom{-}$$0.2$ & $\phantom{-}$$3.2$    \\
   A   &  $\phantom{-}$$1.6$   & $-$$0.2$              & $\phantom{-}$$0.5$  & $\phantom{-}$$0.9$    & $-$$5.7$           & $-$$2.1$ \\
   C   &  $\phantom{-}$$1.4$   & $\phantom{-}$$0.2$    & $\phantom{-}$$6.8$  &           $-$$0.9$    & $-$$1.6$           & $\phantom{-}$$5.9$    \\
   G   &  $\phantom{-}$$1.2$   & $-$$1.3$              & $\phantom{-}$$5.9$  & $\phantom{-}$$0.1$    & $-$$4.6$           & $0$      \\
\end{tabular}
\end{ruledtabular}
\end{table}

  The SO interaction gives rise to a noncollinear magnetic ordering.\cite{Treves}
However, it does not solve the problems of the HF description.
The magnetic GS realized in YTO is $G_{\bf a}$-$A_{\bf b}$-$F_{\bf c}$, which is
consistent with the neutron-scattering data.\cite{Ulrich}
Both spin (${\bf M}_S$) and orbital (${\bf M}_L$) magnetic moments have nonvanishing
projections onto all three orthorhombic axis ${\bf a}$, ${\bf b}$, and
${\bf c}$, which are ordered according with the G-, A-, and F-type, respectively.
The vectors themselves are given by (in $\mu_B$, refereed to the site 1):
${\bf M}_S$$=$($0.05$,$0.83$,$0.34$) and
${\bf M}_L$$=$($-$$0.23$,$-$$0.33$,$0.03$).
The relative weight of the F and A components in this structure
is very sensitive to the value
of $U$. The F component will dominate for smaller $U$, due to the enlarged
K\&K contribution to the OO: e.g.
${\bf M}_S$$=$($-$$0.07$,$-$$0.14$,$0.96$) and
${\bf M}_L$$=$($0.17$,$0.14$,$-$$0.08$) for $U$$=$$2.5$ eV.
The magnetic GS obtained in LTO on the level of HF calculations is
$C_{\bf a}$-$F_{\bf b}$-$A_{\bf c}$, which has large A component along the
${\bf c}$ direction:
${\bf M}_S$$=$($-$$0.13$,$0.18$,$0.89$) and
${\bf M}_L$$=$($-$$0.14$,$-$$0.07$,$-$$0.21$).
The G-type AFM structure is totally excluded from
$C_{\bf a}$-$F_{\bf b}$-$A_{\bf c}$. Therefore, there is a qualitative inconsistency
between results of HF calculations and the experimental data for LTO.
Formally, the problem can be resolved by using larger $U$$=$$4.5$ eV, which enforces
the strong-coupling limit (Table~\ref{tab:SE}) and leads to the new magnetic
GS: $A_{\bf a}$-$G_{\bf b}$-$C_{\bf c}$ with
${\bf M}_S$$=$($0.31$,$0.88$,$-$$0.14$) and ${\bf M}_L$$=$($-$$0.19$,$-$$0.23$,$0.04$).
However, the same $U$ would lead to the new magnetic GS also in
YTO: $C_{\bf a}$-$F_{\bf b}$-$A_{\bf c}$ with
${\bf M}_S$$=$($0.11$,$-$$0.19$,$0.72$) and
${\bf M}_L$$=$($0.09$,$0.12$,$-$$0.09$), in disagreement with the
experiment.\cite{Ulrich}

  In summary, the lattice distortion alone does not provide a coherent explanation
for the unusual magnetic properties of YTO and LTO. The complexity of these compounds
is related with the fact that the CF splitting, the SE and SO interaction energies
are of the same order of magnitude, and should be treated on an equal footing beyond
the mean-field HF approximation.

  I acknowledge communications with V.~I.~Anisimov, A.~Shorikov,
and A.~Mylnikova, who
are working on a similar problem, and A.~I.~Lichtenstein around results of Ref.~\onlinecite{Pavarini}.
I also thank M.~Imada for drawing my attention to Ref.~\onlinecite{Pavarini}, and
stimulating comments. This work was started during my stay
in two previous projects: JRCAT and ERATO-SSS.


\begin{thebibliography}{10}
\bibitem{comment.1}
$J$ is responsible for the first Hund rule. Interactions enforcing the second Hund rule
are neglected in Eq.~(\ref{eqn:energygain}), but
taken into accound in
HF calculations,\cite{PRB03} discuss below.
\bibitem{comment.4}
The TiO$_6$ octahedron in YTO has two long and four short Ti-O bonds.
Thus, the $t_{2g}$ level interacting mainly with the short-distanced
oxygen sites will be split off and shifted to the higher energies due to the
Ti-O hybridization.
\bibitem{comment.5}
Paramaters reported in Ref.~\onlinecite{Mizokawa}
have been recalculated as
$U$$=$$F^0$ and $J$$=$$(F^2$$+$$F^4)/14$, in terms of
Slater's integrals.
\end{thebibliography}
\end{document}